\documentclass[11pt,twoside]{article}

\usepackage{asp2006}
\usepackage{epsf}
\usepackage{psfig}
\usepackage{lscape}

\begin{document}
\title{Insights into Galaxy Evolution from Mid-infrared Wavelengths}   
\author{Ranga-Ram Chary}   
\affil{Spitzer Science Center, Caltech, Pasadena CA 91125}    

\begin{abstract} 
In this paper, I have attempted to highlight key results from deep extragalactic
surveys
at mid-infrared wavelengths. I discuss advances in our understanding of dust
enshrouded star-formation and AGN activity at $0<z<3$ from $IRAS$, $ISO$ 
and $Spitzer$.
The data seem to indicate that about 70\% of the co-moving
star-formation rate density at 0.5$<z<3$ 
is obscured by dust and that AGN, including obscured sources,
 account for $<$20\% of the co-moving bolometric luminosity density. There is tentative
evidence that the mode of star-formation changes as a function of redshift; star-formation
at $z\sim2$ is preferentially in massive, ultraluminous infrared galaxies (ULIRGs)
while $z\sim1$
sources are luminous infrared galaxies (LIRGs) which are about 1 mag fainter than ULIRGs
in the near-infrared. This evolution of the star-formation mode, 
is similar to the evolution of the redshift distribution of X-ray sources as a function of
X-ray luminosity and would suggest an extension of
the downsizing hypothesis to include both AGN and star-forming galaxies.
Measuring dust-enshrouded star-formation at $z>3$ will become possible only
with future facilities like ALMA. Currently, the presence of
dust can only be assessed in a small fraction of the youngest
starbursts at $z>5$ by looking for redshifted large equivalent width
H$\alpha$ emission in broadband filters like the IRAC 4.5$\mu$m
passband. H$\alpha$ to UV ratios in these objects are a tracer of dust extinction 
and measuring this ratio in GOODS galaxies
indicate dust in $\sim$20\% of star-forming galaxies at $z>5$. Finally, implications for reionization 
based on the measured stellar mass density and star-formation rates of galaxies at these redshifts are 
discussed. 

\end{abstract}


\section{Pros and Cons of the Mid-infrared}   

The mid-infrared regime of the electromagnetic spectrum, is generally defined by the wavelength range
which can be detected with arsenic/antimony doped silicon semiconductors and spans 5$-$40$\mu$m. 
Extragalactic surveys in the mid-infrared are typically limited by background noise since the
thermal emission from zodiacal light in the solar system
peaks between 10$-$20$\mu$m \citep{Kelsall:98}. These wavelengths
correspond to the trough 
between the optical/near-infrared and far-infrared wavelengths in the spectrum of the extragalactic background 
light (EBL).
To illustrate, the EBL intensity derived by integrating the light from galaxies detected at
15 and 24$\mu$m is around 2.5-5 nW~m$^{-2}$~sr$^{-1}$ while the EBL intensity
at both 140 $\mu$m and 2.2 $\mu$m
has a value of about $\sim$25 nW~m$^{-2}$~sr$^{-1}$ \citep{Lagache:05}. 
Since the EBL is the sum total of redshifted emission from sources integrated over the observable Universe,
the bulk of which arises from galaxies at $z<1.5$,
this implies that the energy emitted
at near- and far-infrared wavelengths is almost an order of magnitude larger than that at mid-infrared
wavelengths. This is generally true even for individual star-forming galaxies; the mid-infrared regime
lies between the optical/near-infrared stellar photospheric emission and the far-infrared emission from
dust heated to T$\sim$20-40K, both of which together dominate the bolometric luminosity of the galaxy.

The mid-infrared is a complicated spectral region representing the integrated emission
from broad polycyclic aromatic hydrocarbon features, continuum emission from 
dust grains between 300K and 40K and silicate absorption \citep{Desert:90}. 
As a result, even small errors in redshifts translate into large errors in spectral $k-$ corrections making
estimates of the true rest-frame luminosity difficult (Figure 1).

Despite these difficulties, the mid-infrared is the most powerful tracer of dust enshrouded star-formation activity
out to redshifts of 3 (Figure 2). This is possible because of the tight correlation
between mid- and far-infrared emission on both sub-galactic and galactic scales \citep{Calzetti:05, CE01}.
For example, rest-frame 7$\mu$m emission has been shown to trace L$_{\rm IR}$ with
a 1$\sigma$ scatter of 40\% in the local Universe. The correlation appears to hold even
in the $z\sim1$ Universe as determined from a comparison between 7$\mu$m, 70$\mu$m and
1.4 GHz radio luminosities of individual galaxies. However,
there is some evidence that at low
metallicities (Z$<$1/4 Z$_{\sun}$) the PAH line strengths show significant differences \citep{Smith:06}. 
On the instrumentation side, mid-infrared observations have arcsec sized beams even with
small $\sim$1m class telescopes. This has the benefit of alleviating the effects of source confusion. It also 
simplifies the identification of the counterparts of mid-infrared sources
at optical/near-infrared wavelengths and thereby enables the measurement
of their 
spectroscopic redshifts. Finally, mid-infrared arrays
suffer less from latents and transient effects in their
responsivity than far-infrared arrays and
need to be cooled only down to a few degrees Kelvin instead of the milli-kelvin requirements
of submillimeter bolometers. 

\section{Key Extragalactic Mid-infrared Surveys}

Deep $IRAS$ 12 $\mu$m and 60 $\mu$m observations in the direction of the North Ecliptic Pole \citep{Hacking:87},
were the first to provide evidence for redshift evolution of the infrared luminosity function of galaxies.
The survey measured a source density of $\sim$30 deg$^{-2}$
down to 10 mJy which was a factor of 1000 deeper than the $IRAS$ All Sky Survey. The source counts down to this flux
density limit provided tentative evidence for evolution of the infrared luminosity function to redshifts of 0.2 
\citep{HCH:88}.

Subsequent mid-infrared surveys which have revolutionized our
understanding of galaxy evolution at cosmological distances are those undertaken with
$ISOCAM$ at 15 $\mu$m surveys \citep{Genzel:00} and $Spitzer$/MIPS at 24$\mu$m. 

$ISOCAM$ undertook a variety of deep 15$\mu$m surveys, the deepest of which, apart from those in lensing cluster fields, 
was in the Hubble Deep Field North \citep{Aussel:99}. These observations which covered
an area of $\sim$ 20 arcmin$^{2}$ centered on the HDF-N, were sensitive down 
to a flux limit of 100$\mu$Jy and reliably detected about 
$\sim$40 sources. The differential source counts was more than an order of magnitude larger than expected if the local
infrared luminosity function of galaxies did not evolve with redshift. This data provided reliable evidence
for a strong $\sim$(1+z)$^{4}$ luminosity evolution of infrared luminous galaxies. Identification of the 
optical counterparts of the 15$\mu$m sources and measurement of their redshifts from ground-based spectroscopic surveys
\citep{Wirth:04, Cohen:00}
revealed that these were galaxies at $z\sim0.5-1.0$. 
Cross-correlation with X-ray surveys from Chandra and XMM 
(Brandt, these proceedings) as well as their optical spectra
revealed that only about 20\% of these objects appear to have
AGN in them implying that star-formation was powering the bulk of the emission.
As a result, one could apply the local correlation between mid-
and far-infrared luminosities, which
revealed that these were infrared luminous galaxies
with L$_{IR}$=L(8$-$1000$\mu$m)$>10^{11}$~L$_{\sun}$ with the majority of them being LIRGs (10$^{11}<$~L$_{IR}<10^{12}$~L$_{\sun}$).
The estimates of their star-formation rates are in the range of 10-100 M$_{\sun}$~yr$^{-1}$, most of which
is dust obscured. For the subset of sources which are radio detected, the
derived far-infrared luminosities of these sources agree well with
the radio-FIR correlation in the local Universe providing
evidence against an evolution 
of the far-infrared spectral energy distribution with redshift.
The SED fitting also yielded a contribution of $ISOCAM$ galaxies to the
DIRBE observed far-infrared background of 65$\pm$30\% \citep{Elbaz:02}.
The contribution of dusty starbursts to the total EBL has since been
confirmed more recently by stacking $Spitzer$ far-infrared
data on mid-infrared sources by \citet{Dole:06}.

Phenomenological models which attempted to fit the source counts from surveys at various wavelengths, based on 
local galaxy templates, indicated a strong evolution of the far-infrared luminosity density with redshift \citep{CE01, Franceschini:01}.
These models revealed that the far-infrared luminosity density is between a factor of 5$-$10 higher than the
UV luminosity density at $z\sim1$ implying an increase in the average dust extinction in the Universe from
E(B-V)$\sim$0.15 at $z\sim1$ to E(B-V)$\sim$0.3 at $z\sim1$.
However, it is difficult to have these models account for 
cosmic variance which cause source counts to vary by as much as a factor
of 2 between different fields. Also, phenomenological models
are degenerate in factoring in the amount of luminosity evolution and density
evolution in the infrared luminosity function due to the difficulty in
measuring the faint end slope of the luminosity function. Various 
models which simultaneously fit
the source counts at different wavelengths have slightly different 
parameters for the redshift evolution of LIRGs and 
ULIRGs \citep{CE01, Lagache:04}. In addition,
the underlying mid and far-infrared templates used in the models
strongly affect the evolution parameters. Nevertheless, the evolution of the
far-infrared luminosity density appears to be robust to these parameters.

There were two significant problems.
Due to the fact that the PAH features get redshifted out of the $ISOCAM$ 15$\mu$m bandpass at $z\sim1.2$,
it was impossible to assess evidence for evolution of the infrared
luminosity function at higher redshift without using galaxy
counts at 850$\mu$m which are 
discussed elsewhere (R. Ivison, these proceedings).
Secondly, the integral of the co-moving star-formation rate density from the
models overpredicted the measured 
comoving stellar mass density seen by \citet{Dickinson:03}. This could either
be due to the stellar mass density being underestimated from the uncertain
contribution of the high Mass/Light ratio old stellar component in galaxies. 
Alternately,
it could be due to the fact that at high redshifts, the mid-infrared determined
bolometric luminosity densities are overestimates. Deep, wide $Spitzer$ surveys
(e.g. Dickinson et al. $Spitzer$ GO-3) are well positioned 
to address this problem by providing 70$\mu$m coverage of large 
numbers of $z>2$ galaxies, enabling bolometric luminosities to be validated
using stacking analysis.

\section{GOODS MIPS 24$\mu$m Survey}

The Great Observatories Origins Deep Survey 24$\mu$m observations were taken with the Multiband 
Imaging Photometer Spectrometer (MIPS; \citet{Rieke:04}). The 24$\mu$m passband can detect PAH emission out
to $z\sim3$. However, unless the observations are adequately deep, observations would be unable to detect typical
infrared luminous galaxies much beyond the limits set by $ISOCAM$. For example, at a flux limit of 80$\mu$Jy, 24 micron
observations are sensitive to a typical 3$\times$10$^{11}$~L$_{\sun}$ out to $z\sim1.2$ which is about the 
same as what the $ISOCAM$ HDF observations could detect such objects out to. By going down to 20$\mu$Jy, such 
galaxies are now being detected out to redshifts 2.5, enabling a comparison with the $z\sim2$ submillimeter
galaxy population and $z\sim3$ Lyman-break galaxy population.

The $Spitzer$ beam size at 24$\mu$m is 5.7$\arcsec$ FWHM which raises the issue of source confusion. However, due to the accurate
pointing of $Spitzer$ and $\sim$0.2$\arcsec$ rms with which we can align the 24$\mu$m images with 
observations at 3$-$8$\mu$m,
the concerns of confusion noise placing a floor on the sensitivity are misplaced. 
The surface density of sources is 17 arcmin$^{-2}$ in GOODS-N and 24 arcmin$^{-2}$ in GOODS-S.
Also, at this resolution, the vast majority of extragalactic sources are point sources.
Using a prior position based point source fitting algorithm, we have been able to extract sources with a
completeness of 84\% at 24$\mu$Jy (Figure 3). The 5$\sigma$ point source sensitivity limit of this survey is 25$\mu$Jy. The fundamental reason for the success of this technique is because every MIPS source
is detected in deep 3.6$\mu$m observations. 

The differential 24$\mu$m source counts are similar in shape to the 15$\mu$m counts
with a break in dN/dS at 400$\mu$Jy (Figure 3). The slope of the counts at fainter fluxes is $\sim$S$^{-1.6}$.
One of the most salient features of the 24$\mu$m counts is they peak (in dN/dS$\times$S$^{2.5}$)
at fainter flux values than predicted from the phenomenological models which are fit to
the multiwavelength counts. Attempts to resolve this have included changing the evolution
of the LIRG to ULIRG ratio with redshift as in \citet{Chary:04} or weakening the strength
of the 11 and 12$\mu$m PAH \citep{Lagache:04} in the galaxy templates so that objects which contribute to the peak of
the 15$\mu$m counts contribute at fainter fluxes to the 24$\mu$m counts.

There is little observational evidence for a weakening of the 11 and 12$\mu$m PAH (e.g. Pope et al. 2006, these proceedings). Stacking 
of IRS spectra of 24$\mu$m sources indicate
that apart from an underlying hot dust continuum which might lower
the equivalent widths of all PAH between 6 and 12$\mu$m, the relative line ratios 
do not vary except for metal poor galaxies. Since dusty starbursts appear to be in massive 
galaxies, which have also enriched the ISM in the process of increasing the dust content,
there is less concern that the bolometric luminosities for the bulk of the galaxies are 
incorrect although some recalibration of the libraries probably have to be done for the SEDs
of the most luminous galaxies (L$_{IR}>10^{13}$~L$_{\sun}$)
for which no low-redshift counterparts exist.

\section{Evidence for ``Downsizing'' in AGN and Galaxies}

From the redshifts and 24$\mu$m flux densities of individual sources, we can calculate
their far-infrared luminosity using the template 
fitting approach \citep{CE01}. 
The accuracy of this approach has been verified at $z<2$ from 70$\mu$m
and 850$\mu$m stacking of LIRGs and ULIRGs. Also, the spectral
energy distribution of galaxies at $z\sim2$ selected using the BzK technique
also appear to match the templates spectral energy distribution closely \citep{Daddi:05}. The far-infrared
luminosity can be converted into a star-formation rate using the conversion factor
in \citet{Kennicutt:98}. One possible uncertainty in this conversion is
estimating the contribution to the far-infrared luminosity
from dust being heated by the evolved stellar population. Deep optical surveys
provide a measure of the star-formation seen in the rest-frame ultraviolet. Since the bulk
of the far-infrared emission is dust heated by the ultraviolet 
radiation field which is dominated by star-formation, adding the
UV and FIR star-formation rates yields the star-formation 
rates of individual objects.

Although the stellar masses of galaxies can be derived using SED fits to the multiband
photometry, a comparison between the 3.6$\mu$m magnitudes of LIRGs and ULIRGs which traces
rest-frame near-infrared emission provides a good estimate of the stellar mass.
The photometry indicates
that LIRGs at $z\sim1.5$ have a median 3.6$\mu$m brightness 
of 21.2 AB mag while ULIRGs are 20.5 AB mag. At $z\sim2.4$, LIRGs 
are 22.5 AB mag while ULIRGs are 21.5 AB mag. Thus ULIRGs are systematically,
about 1 AB mag brighter at 3.6$\mu$m indicating that they are more massive.

Finally, AGN can be estimated in the $0.5<z<3$ range either by identifying power laws from
hot dust in the observed 5.8$-$24$\mu$m range. Alternately, the X-ray luminosity and 
photon index of sources provides a good measure of AGN activity. We caution that 
Compton-thick AGN are undetected even in the deepest Chandra surveys beyond $z\sim0.6$ and
the IRAC power law identification of AGN breaks down at redshifts of $z>2$ because dust sublimes
at around $\sim$1500$-$2000~K corresponding to rest-frame $\sim$2$\mu$m. At $z\sim2$, the 2$\mu$m
emitting hot dust is redshifted out of the 5.8$\mu$m bandpass.

Putting these various datasets together, we confirm the ISOCAM result
implying an evolution of the far-infrared
luminosity density with redshift \citep{Babbedge:06, lef:05, pablo:05}.
However, the data also indicate an evolution of the galaxies dominating the co-moving star-formation rate
density. At $z\sim1$, LIRGs which typically have stellar masses of $\sim$10$^{9-10}$~M$_{\sun}$
and star-formation rates of 10-100~M$_{\sun}$~yr$^{-1}$ are dominating the energetics. At
$z\sim2$, it is the ULIRGs which have stellar masses $>$10$^{10}$~M$_{\sun}$ that are 
dominating the star-formation rate density (Figure 4). The AGN contribution to the luminosity density
at both redshifts appear to less than $\sim$20\%. Thus, lower mass galaxies form later
in cosmic time and more massive galaxies build their stellar mass earlier.

This trend is similar to the redshift distribution of X-ray sources as a function of X-ray luminosity
\citep{Ueda:03}. The X-ray data seem to suggest that the most massive ($\sim$10$^{9}$~M$_{\sun}$)
nuclear black holes
built up their mass in a luminous accretion event around $z\sim2-3$ while the ULIRGs were building
up their stellar mass. The lower mass black holes on the other hand, grew around $z\sim1$
when the LIRGs were building up their stellar mass providing
impetus for the downsizing scenario extending to AGN and starbursts. Since low mass galaxies are more numerous
and populate the Hubble sequence, this implies that the Hubble sequence
was probably in place only at $z<1$.

However, there are potential systematics. There is some evidence in our IRS
spectra that the color temperatures of the mid-infrared spectra are relatively cool.
Templates of local galaxies which are used to derive the bolometric corrections typically
have steeply rising hot dust continuum with increasing L$_{IR}$. This would imply that
bolometric corrections might have been overestimated and many of the objects are less luminous.
The effect is particularly important in the $2<z<3$ regime where we have limited checks
on the shape of the template from stacking analysis. There is also evidence in some objects
that continuum from an AGN results in
the 6.2$\mu$m PAH equivalent width being reduced by a factor of 3 compared to
local starburst galaxies such as M82. This would again imply an incorrect bolometric correction
in the $z>2$ sources. Finally, due to the difficulty in identifying AGN at $z>2$ from the IRAC colors,
we might be underestimating the AGN contribution which would be predominantly to the brightest
24$\mu$m sources thereby lowering the ULIRG contribution. Further checks, especially in the $2<z<3$
range, where both photometric and spectroscopic redshifts are limited are essential to 
verify the result shown in Figure 4.

\section{Dust at Higher Redshifts and Implications for Reionization}

Detecting dust emission at redshifts greater than 3 is difficult. 
The strongest PAH features at 6.2 and 7.7 $\mu$m are redshifted out of the MIPS 24$\mu$m bandpass. 
Only source with large AGN contributions displaying significant hot dust emission are detected at $z>3$ at 24$\mu$m. 
It 
appears that the equivalent width of the 3.3$\mu$m PAH is low enough that it would be undetectable above 
the stellar continuum and  the correlation between 3.3$\mu$m emission 
and L$_{\rm IR}$ is not well established.
70$\mu$m surveys are about one to two orders of magnitude less sensitive than required to detect dust in all 
but the most hyperluminous sources. As a result, at $z>3$, 850$\mu$m surveys are best suited for detecting dust 
emission but the current sensitivity limits (Figure 2) will not be significantly improved until 
the Atacama Large Millimeter Array (ALMA) becomes operational.

One possible technique has been to measure redshifted H$\alpha$ emission through broadband photometry.
This technique can only be applied over a limited redshift range. 
In \citet{CSE:05}, we demonstrated the possibility that the enhanced 4.5$\mu$m/3.6$\mu$m flux ratio
of the lensed $z=6.56$ galaxy
HCM6A is due to large equivalent width H$\alpha$. A study of the broadband photometry in GOODS galaxies has 
since revealed that about 10$-$20\% of spectroscopically confirmed $z>5$ objects show a similar enhancement in the 4.5$\mu$m
flux \citep[Figure 5;][]{Chary:06}. The H$\alpha$ determined star-formation rates are on average higher by a factor of $\sim$3, compared to the UV determined
star-formation rate, indicating the presence of dust in a small fraction of high redshift objects. The presence
of dust within 1 Gyr of the Big Bang argues for the production of dust from SNe since red giants and AGB stars have had
insufficient time to evolve.

Since the dust content correlates with the number of SNe, it must increase with decreasing redshift and is presumably directly 
correlated to the increase in stellar mass.
This implies that for a small fraction of objects, the measured star-formation rate from the UV continuum is lower
than the true star-formation rate. Since typical starbursts have an e-folding time associated with them, the conclusion
is that that the star-formation rate ($\dot{\rho}_{*}(t)$) of a galaxy at some time $t<t'$ in its past history must be: 

\begin{equation}
\dot{\rho}_{*}(t) = \dot{\rho}_{*}(t') \exp[-(t-t')/\tau_{SFR}] \exp[-\tau(t)+\tau(t')]
\end{equation}

where,
\begin{equation}
\tau(t') \propto \int_{0}^{t'} \dot{\rho}_{*}(t) dt
\end{equation}

$\tau_{SFR}$ is the e-folding time for the star-burst while $\tau(t)$ is the extinction in the star-formation
rate metric (i.e. UV wavelengths) at time $t$. Naturally, $\tau(t)<\tau(t')$.

We have now been able to measure the star-formation rate and stellar mass density in galaxies at $z\sim6$. For
canonical estimates of the escape fraction (fesc$\sim$0.15) and clumping factor (C=30), the measured
star-formation rate at $z>6$ falls more than a factor of 10 short of the minimum needed to reionize the IGM. However, if the dust
content is increasing with redshift, the past star-formation rate must be higher not just by the e-folding time
of the burst but also by the increase in dust extinction which is directly proportional to the stellar mass.

The past history of star-formation can either be identified by fitting SEDs to the multiband photometry and
extrapolating backwards in time or
by fitting an evolving
starburst model to the measured
star-formation rate and stellar mass density at $z\sim6$ (Figure 6). 
From rest-frame ultraviolet observations of
Lyman-break galaxy samples \citep{Bouwens:06}, it appears that the 
star-formation rate density is decreasing at $z>6$, while
from the extrapolation process the star-formation rate density must be
increasing at higher redshifts to reproduce the observed stellar mass density. 
This discrepancy is probably attributable to the large completeness corrections
associated with identifying galaxies at $z>7$ using current instrumentation.

The primary conclusions are as follows: 1) If the measured star-formation rate
density and stellar mass density at $z\sim6$ are accurate \citep{Yan:06, Chary:06, Eyles:05, Stark:06}, 
then reionization must be a
brief, inhomogeneous processing lasting only 20-50 Myr at $6<z<7$;
2) If the process of reionization starts at $z\sim9$ and is prolonged,
then the stars that dominate the reionization process must not contribute
to the stellar mass density at $z\sim6$ i.e. the mass of stars and remnants
at $z\sim6$ must be higher than the measured stellar mass by a factor of $\sim$10. This would suggest a contribution from high mass, zero metallicity 
Population III stars or from the low mass end of the galaxy mass function.
Either of these scenarios are unnecessary
if the escape fraction and clumping factor
are significantly different from the canonical estimates of 0.15 and 30 respectively.


\acknowledgements 

I wish to thank the organizers for a very productive and enjoyable meeting.
I would like to acknowledge contributions from various members of the GOODS team which resulted
in the material presented in this paper. This work is partly funded by NASA through Contract 
Number 1224666 issued by JPL/Caltech under NASA contract 1407. 


\clearpage

\begin{figure}
\plotfiddle{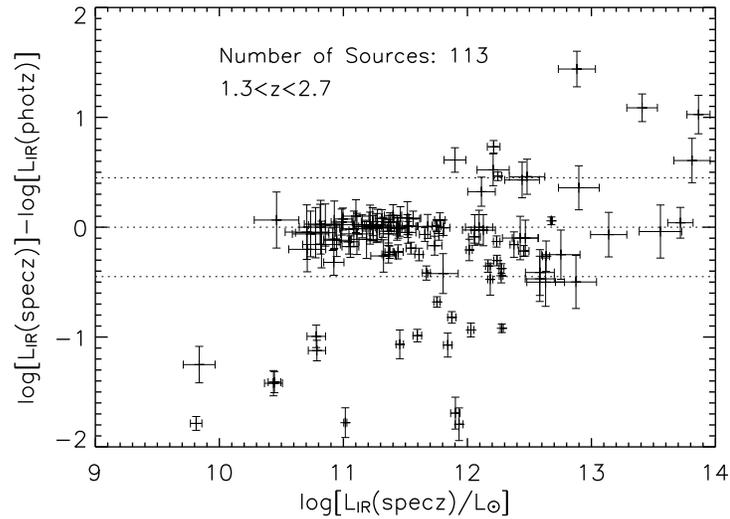}{2.5in}{0}{60}{60}{-170}{-220}
\caption{Uncertainty in determining L$_{\rm IR}$ from monochromatic observed 24$\mu$m fluxes due to an error in the
photometric redshift estimate. Due to the interplay of emission and absorption features in the mid-infrared, small errors
in redshift estimates result in the incorrect far-infrared template being chosen from template libraries.
This manifests itself into large errors in the bolometric correction and thereby the derived bolometric
luminosity of objects. The 1$\sigma$ scatter in the derived L$_{\rm IR}$ due to typical optical/near-infrared
photometric redshifts is a factor of 3 for a 1$\sigma$ scatter in $\Delta$z of 0.24 which corresponds to a $\sigma$($\Delta$z/(1+z$_{spec}$)$\sim$0.1.
}
\end{figure}

\begin{figure}
\plotfiddle{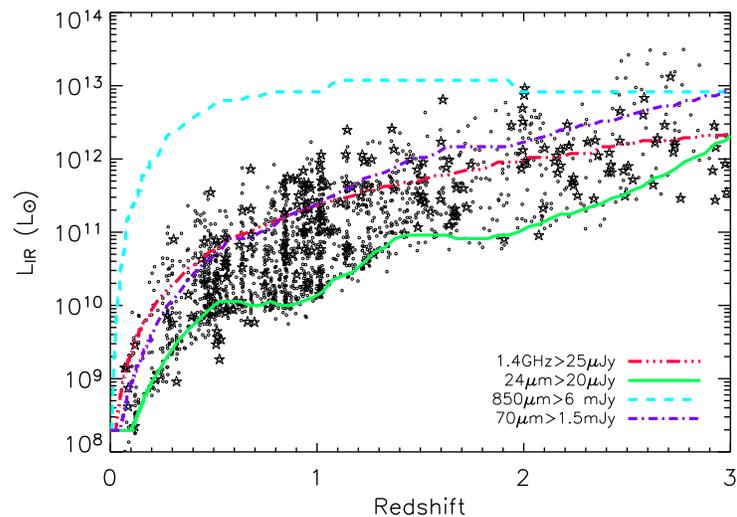}{2.5in}{0}{60}{60}{-170}{-220}
\caption{Sensitivity of deep surveys at different wavelengths
to L$_{\rm IR}$=L(8$-$1000~$\mu$m), a proxy for star-formation rate. 
Also plotted are sources with either photometric or spectroscopic redshifts.
Stars indicate X-ray sources i.e. AGN, while circles are star-forming galaxies.
The GOODS 24~$\mu$m observations are at least an order of magnitude more sensitive
to dust obscured star-formation and AGN activity than other surveys.
}
\end{figure}

\begin{figure}
\plotfiddle{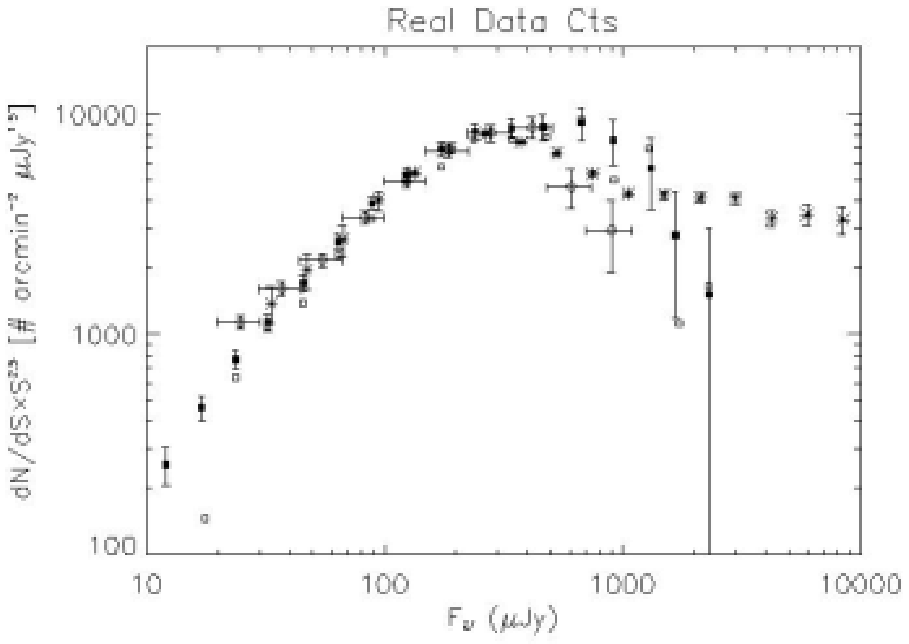}{2.5in}{0}{110}{110}{-160}{-10}
\caption{Differential extragalactic source counts at 24$\mu$m in the GOODS-N field. 
The empty squares are the measured counts, the solid squares are the counts corrected for completeness,
the open diamonds are the completeness corrected counts in the GOODS test field in ELAIS-N1 \citep{Chary:04},
while the asterisks are the counts in various other fields \citep{Papovich:04, Marleau:04}.
Application of the prior-based deblending
technique results in an extracted source catalog which is 84\% complete at 24$\mu$Jy illustrating that
source confusion is not a significant source of noise in GOODS. 
}
\end{figure}

\begin{figure}
\plotfiddle{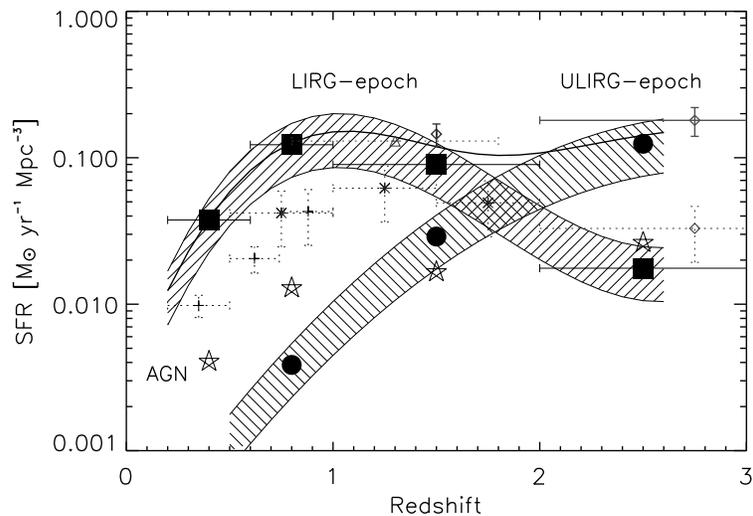}{2.5in}{0}{60}{60}{-170}{-220}
\caption{Relative contribution of LIRGs (hatched region), ULIRGs (backhatched region)
and AGN (stars) to the co-moving far-infrared luminosity density between $0<z<3$. The dotted
symbols underlying are various measures of star-formation
from the UV and H$\alpha$ referenced in \citet{CE01}.
The systematic uncertainty in this measurement is a factor of $3-4$ at $z>2$ due to the limited
number of sources with spectroscopic redshifts and uncertainty in the AGN contribution (See text). 
}
\end{figure}

\begin{figure}
\plotfiddle{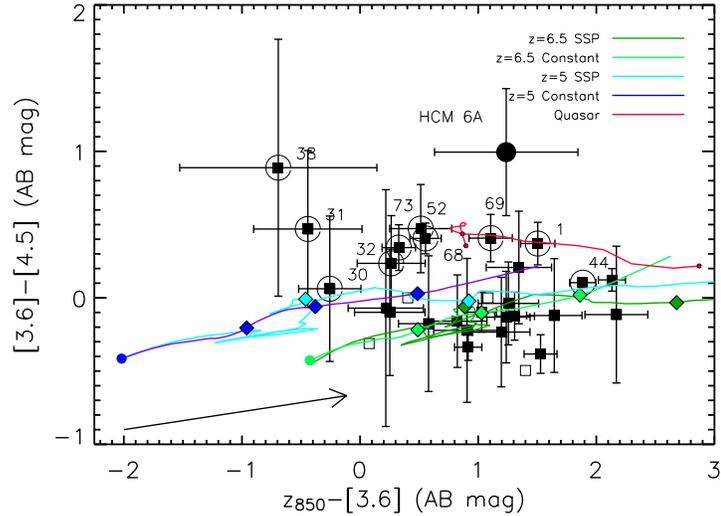}{2.5in}{0}{60}{60}{-170}{-220}
\caption{Identification of H$\alpha$ in emission using broadband colors. Circled sources
are candidate H$\alpha$ emitters with $z_{spec}>5$ \citep{Chary:06, CSE:05}. Comparison of the H$\alpha$
SFR to the UV continuum derived star-formation rate provides a unique technique to measure dust obscured
star-formation at $z>5$. However, this technique can be used only in the most violently star-forming
galaxies because the equivalent width of H$\alpha$ must be very high for it to be detectable in a 
broadband filter like $Spitzer$/IRAC 4.5$\mu$m. 
}
\end{figure}

\begin{figure}
\plotfiddle{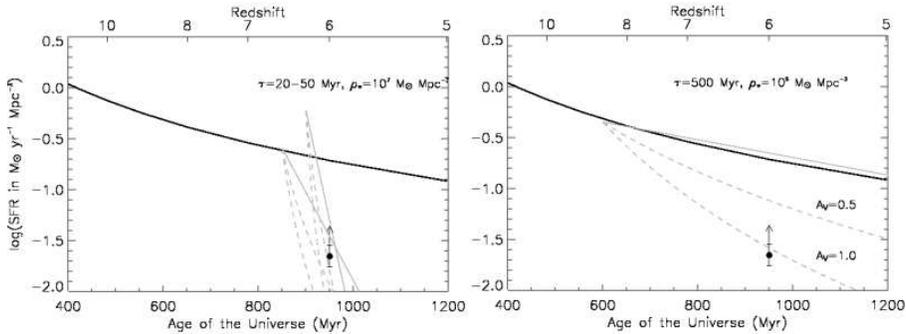}{1.5in}{0}{100}{100}{-180}{-10}
\caption{
Figure 1: (Left Panel) The solid black line shows the minimum star-formation 
rate as a function of redshift, for a Salpeter IMF, which is required to reionize the IGM (escape fraction of 0.15 and clumping factor of 30). The black symbol shows the measured star-formation rate density at z$\sim$6 (Bouwens et al. 2006). The solid grey line, shows the star-formation history of the Universe required to reproduce the measured stellar mass density as well as the star-formation rate density at z$\sim$6. Since the grey line can exceed the solid line for a brief interval of time, reionization must be a brief process lasting 20-50 Myr at z$\sim$6.5. (Right Panel) If on the other hand reionization took place at z$\sim$9, then either the stellar mass density at redshift 6 must be $\times$10 higher than what is measured due to a steep faint end mass function or the sources responsible for reionization at z$\sim$9 do not contribute to the stellar mass density at z$\sim$6. The dashed grey lines show the role of increasing
extinction on the measured star-formation rate density for A$_{\rm V}$=0.5 mag and A$_{\rm V}$ of 1 mag. 
}
\end{figure}

\end{document}